\documentclass[sigconf]{acmart}

\usepackage{url}
\usepackage{graphicx}
\usepackage{amsmath}
\usepackage{amsthm}
\usepackage{booktabs}
\usepackage{algorithm}
\usepackage{algorithmic}
\usepackage{color}
\usepackage{xspace}
\usepackage{acmart-taps}
\usepackage[misc]{ifsym}

\usepackage[ruled, vlined, nofillcomment, linesnumbered, algo2e]{algorithm2e}

\usepackage{booktabs}
\usepackage{multirow}
\usepackage{balance}

\newcommand{\aka}{\emph{a.k.a.,}\xspace}
\newcommand{\eg}{\emph{e.g.,}\xspace}

\newcommand{\ie}{\emph{i.e.,}\xspace}
\newcommand{\ignore}[1]{}
\newcommand{\paratitle}[1]{\vspace{1.5ex}\noindent\textbf{#1}}

\usepackage{bm}

\AtBeginDocument{%
  }

\setcopyright{acmcopyright}
\copyrightyear{2022}
\acmYear{2022}
\acmDOI{10.1145/3523227.3546752}
\acmConference[RecSys '22]{Proceedings of the 16th ACM Conference on Recommender Systems}{September 18--23, 2022}{Seattle, WA, USA}
\acmBooktitle{Proceedings of the 16th ACM Conference on Recommender Systems (RecSys '22), September 18--23, 2022, Seattle, WA, USA}
\acmPrice{15.00}
\acmISBN{978-1-4503-XXXX-0/22/xx}
\begin{document}


\title{Modeling Two-Way Selection Preference for Person-Job Fit}

\author{Chen Yang$^{\dagger}$}
\email{2021100973@ruc.edu.cn}
\affiliation{
         \institution{Gaoling School of Artificial Intelligence, Renmin University of China}
      \city{Beijing}
  \postcode{100872}
    \country{China}
}

\author{Yupeng Hou$^{\dagger}$}
\email{houyupeng@ruc.edu.cn}
\affiliation{
         \institution{Gaoling School of Artificial Intelligence, Renmin University of China}
      \city{Beijing}
  \postcode{100872}
    \country{China}
}

\author{Yang Song}
\email{songyang@kanzhun.com}
\affiliation{%
  \institution{BOSS Zhipin}
  \city{Beijing}
  \country{China}
}

\author{Tao Zhang}
\email{kylen.zhang@kanzhun.com}
\affiliation{%
  \institution{BOSS Zhipin}
  \city{Beijing}
  \country{China}
}

\author{Ji-Rong Wen$^{\dagger}$}
\email{jrwen@ruc.edu.cn}
\affiliation{
    \institution{Gaoling School of Artificial Intelligence, Renmin University of China}
  \city{Beijing}
  \postcode{100872}
  \country{China}
}

\author{Wayne Xin Zhao$^{\dagger\ddagger}$
\textsuperscript{\Letter}
}
\email{batmanfly@gmail.com}
\orcid{1234-5678-9012-3456}
\affiliation{
    \institution{Gaoling School of Artificial Intelligence, Renmin University of China}
  \city{Beijing}
  \postcode{100872}
  \country{China}
}

\thanks{$\dagger$ Beijing Key Laboratory of Big Data Management and Analysis Methods.}
\thanks{$\ddagger$ Beijing Academy of Artificial Intelligence, Beijing, 100084, China.}
\thanks{\includegraphics[]{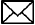} Corresponding author.}

\renewcommand{\authors}{Chen Yang, Yupeng Hou, Yang Song, Tao Zhang, Ji-Rong Wen, Wayne Xin Zhao }
\renewcommand{\shortauthors}{Chen Yang et al.}

\begin{abstract}

Person-job fit is the core technique of online recruitment platforms, which can improve the efficiency of recruitment by accurately matching the job positions with the job seekers. 
Existing works mainly focus on modeling the unidirectional process or overall matching.
However, recruitment is a two-way selection process, which means that both candidate and employer involved in the interaction should meet the expectation of each other, instead of unilateral satisfaction. 
In this paper, we propose a dual-perspective graph representation learning approach to model directed interactions between candidates and jobs.
To model the two-way selection preference from the dual-perspective of job seekers and employers,
we incorporate two different nodes for each candidate (or job) and characterize both successful matching and failed matching via a unified dual-perspective interaction graph.
To learn dual-perspective node representations effectively, we design an effective optimization algorithm, which involves a quadruple-based loss and a dual-perspective contrastive learning loss.
Extensive experiments on three large real-world recruitment datasets have shown the effectiveness of our approach. Our code is available at \textcolor{blue}{\url{https://github.com/RUCAIBox/DPGNN}}.

\end{abstract}

\ccsdesc[500]{Information systems~Recommender systems}

\keywords{person-job fit, graph neural network, contrastive learning}

\maketitle

\section{Introduction}

With the rapid development of Internet technology, online recruitment has become a prevalent service for job hunting, which matches qualified candidates with suitable jobs.
Due to the massive growth of candidates and recruiters in online recruitment, it is critical to design effective algorithms to automatically establish  high-quality connections between job seekers and employers through the recommendation mechanism.
Such a task is called \emph{person-job fit}~\cite{zhu2018person}.

In the literature, a variety of studies have been conducted for person-job fit, such as reciprocal recommendation~\cite{malinowski2006matching}, job recommendation~\cite{lee2007fighting, paparrizos2011machine, zhang2014research} and job-oriented skill measurement~\cite{xu2018measuring}. 
Since the information of job and candidate is usually organized and described in textual documents, many text-based person-job fit works have been proposed in recent years, which learn matching function by modeling the textual semantics~\cite{zhu2018person,qin2018enhancing}. These methods might also be thought of as content-based reciprocal recommendation approaches.

Despite the performance improvement, existing studies either model the one-way selection process (\eg recommending qualified candidates conditioned on the job requirement~\cite{ye2019identifying,su2022optimizing}) or overall matching relation (\eg the text matching approach~\cite{zhu2018person}). 
Since the recruitment process involves the two sides of candidates and employers, it naturally reflects the two-way selection preference~\cite{mcnamara1990job} from both perspectives of candidates and job positions. 
In order to achieve a person-job matching, both sides involved in the interaction should meet the expectation of each other, instead of unilateral satisfaction. 
Such a kind of bilateral satisfaction is the key to the success of person-job fit.
On online recruitment platforms, it is common to see unilateral satisfaction cases. For example, a job seeker applies for a position but is refused by the employer, or an employer sends an interview request but is refused by the job seeker. These unilateral interactions correspond to failed matches, which don't meet the bilateral expectation. 
Actually, two-way selection is a common phenomenon in nature and society~\cite{zhou2014model}, and it is essential to consider such a bilateral match process in person-job fit. 

To improve the person-job fit, we explicitly model the two-way selection preference from the  dual perspectives of  job seekers and employers.
We develop our solution based on a graph representation learning approach.  
The core idea is to incorporate two different nodes for each candidate (or job): one captures its own preference to select a job (or a candidate) and the other reflects the corresponding preference from others. In this way, we can characterize both successful matching and failed matching via a unified dual-perspective interaction graph.   
Based on such an interaction graph,
we propose a novel dual-perspective graph representation learning approach for modeling two-way selection intentions. There are two major technical contributions in our approach. Firstly, we learn both active and passive representations via a hybrid preference propagation, capturing the node  characteristics to actively select and to be passively selected for person-job interaction. Secondly, we design an effective optimization 
algorithm for learning dual-perspective node representations, which involves a quadruple-based loss and a dual-perspective contrastive learning loss.

To the best of our knowledge, it is the first study
    that explicitly models the two-way selection preference for person-job fit. In order to demonstrate the effectiveness of our approach, 
    we conduct extensive experiments on real-world recruitment data from three job domains. Experiment results show that our approach is significantly better than both collaborative filtering baselines and content-based baselines.

\section{Related Work}

\subsection{Person-Job Fit}
As an important task in recruitment data mining~\cite{kenthapadi2017personalized,shalaby2017help}, Person-Job Fit (PJF) has been widely studied.
The early work can be traced back to~\cite{malinowski2006matching}, in which expectation maximization algorithm was used to make recommendations 
by utilizing profile information from both candidates and jobs. 
Some other works tackled this challenge from the perspective of collaborative filtering~\cite{diaby2013toward,lu2013recommender}. Zhang \textit{et al.}~\shortcite{zhang2014research} systematically compared a number of user-based and item-based collaborative filtering algorithms in order to recommend suitable jobs for candidates. 

In recent years, most research cast this problem as a text matching task in order to make full use of the rich textual semantic information in resumes and job requirements. 
Some works have been proposed to utilize various kinds of neural networks to represent resumes and job posts, such as
CNN~\cite{zhu2018person,shen2018joint}, RNN~\cite{qin2018enhancing} and memory network~\cite{yan2019interview}.
Besides, some researchers have explored various techniques to improve the expression ability of the text encoders, such as adversarial learning~\cite{luo2019resumegan}, transfer learning~\cite{bian2019domain} and co-teaching mechanisms~\cite{bian2020learning}.

Some works also considered the multi-granularity interactions in the person-job fit scenario.
Le~\shortcite{le2019towards} cast the multi-level interactions as supervision signals and proposed a ranking-based loss function.
Fu~\shortcite{fu2021beyond} proposed a bilateral multi-behavior sequence model to describe the dynamic comprehensive preferences of users. 
Besides, some studies have explored other related behaviors of users to improve person-job fit, such as search history~\cite{hou2022leveraging} and interactive feedbacks~\cite{fu2022market}.
However, the two-way selection preference, which naturally exists in online recruitment scenarios, is not explicitly modeled in these methods. In this work, we propose a novel dual-perspective graph representation learning approach that jointly leverages interactions and text.

\subsection{Recommender Systems}
Recommender systems have been widely deployed to alleviate information overload on the web. 
A prevalent technique in modern recommender systems is collaborative filtering (CF)~\cite{he2017neural,he2020lightgcn}. The core idea is to predict users' preferences by exploiting their historical interactions. 
Various collaborative filtering methods have been proposed, such as matrix factorization-based methods~\cite{rendle2009bprmf}, neural network-based methods~\cite{he2017neural} and auto-encoder-based methods~\cite{liang2018multivae}.

Recently, graph-based collaborative filtering (GCF) is proposed to deepen the use of high-hop neighbors of users and items~\cite{wang2019neural,he2020lightgcn}.
These methods usually organize interaction data into a bipartite graph. However, in the person-job fit scenario, it's difficult to leverage directional interactions and model the two-way selection preference via existing GCF methods. Thus, in this work, we 
redesign the network architecture and 
incorporate two different nodes for each candidate (or job) and propose a graph collaborative approach on the constructed dual-perspective interaction graph.

Since recruitment is a two-way selection process, it is also related to a research topic called Reciprocal Recommender System (RRS)~\cite{siting2012job, mine2013reciprocal, yu2011reciprocal}, a kind of recommender system that recommends users to other users rather than items.
RRS is frequently employed in domains with significant societal influences, such as recruitment~\cite{mine2013reciprocal, yu2011reciprocal, su2022optimizing},  online dating~\cite{pizzato2010recon, xia2015reciprocal, tu2014online} and social networking platforms~\cite{he2010social}.
They can be divided into two main categories based on the data source they used: content-based methods based on user profiles~\cite{pizzato2010recon, neve2020imrec} and collaborative filtering methods based on user behaviors~\cite{xia2015reciprocal, kleinerman2018optimally, neve2019latent}.
As mentioned above, person-job fit methods based on text matching~\cite{zhu2018person,yan2019interview} can also be regarded as content-based reciprocal recommender systems.
However, existing RRS methods do not model directional behaviors to explicitly capture two-way selection preferences, which is important in person-job fit scenario~\cite{mcnamara1990job}. In this work, we first construct a dual-perspective interaction graph based on directional interactions, and then propose a hybrid method to jointly model the interactions and text via graph neural networks.

\section{Proposed Method}
\begin{figure*}[t!]
	\centering
	\includegraphics[width=0.8\textwidth]{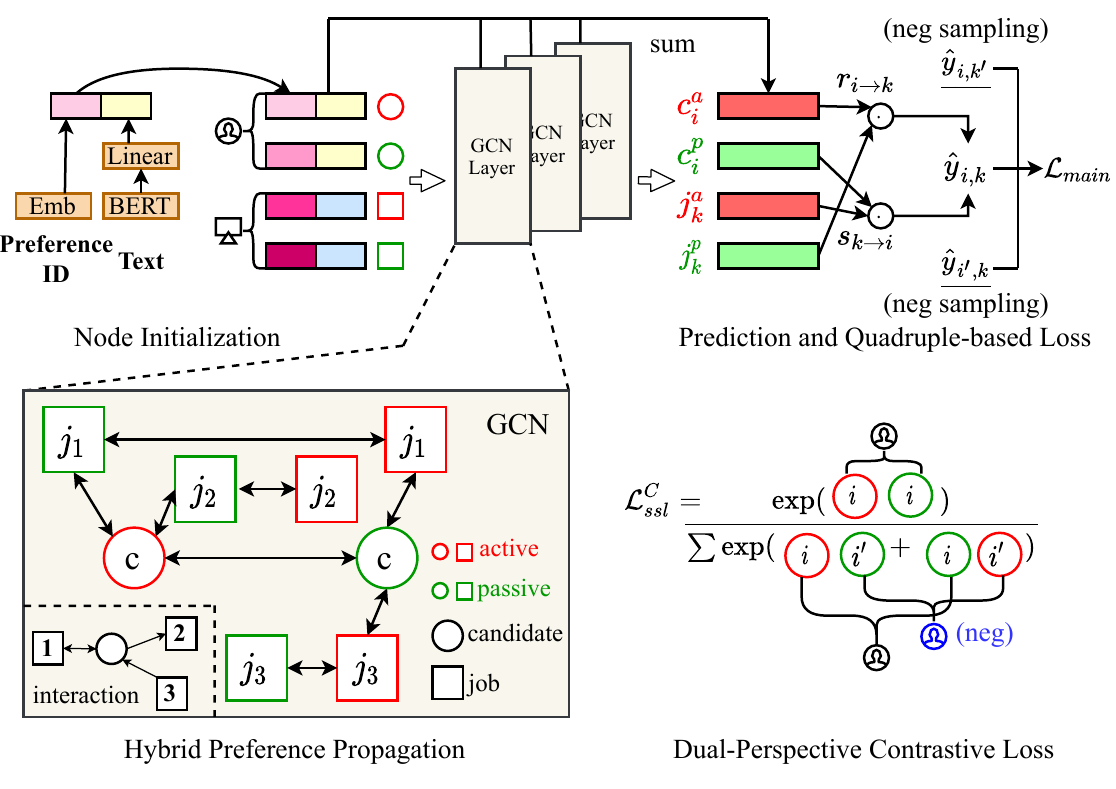}
	\caption{Overall framework of the proposed model DPGNN. 
	Note that in the hybrid preference propagation module (Sec.~\ref{sec:dpg_construct}), 
	we construct a new dual perspective interaction graph based on original interactions. Each candidate (job) corresponds to two nodes in the constructed graph, representing her active and passive preferences.
	Edges are created between nodes that belong to the same user as well as between nodes whose corresponding users are connected by directional interactions.
	}
	\label{fig:model}
\end{figure*}

In this section, we first formulate the person-job fit task and then introduce the proposed \textbf{D}ual-\textbf{P}erspective \textbf{G}raph \textbf{N}eural \textbf{N}etwork, named \textbf{DPGNN}.
The overall architecture is depicted in Figure~\ref{fig:model}.

\subsection{Notations and Problem Statement}\label{sec:not}
Assume that we have a set of candidates $\mathcal{C} = \{c_1, c_2,...,c_{n}\}$ and a set of jobs $\mathcal{J} = \{j_1, j_2,...,j_{m}\}$, where $n$ and $m$ are the total numbers of candidates and jobs. 
Each candidate or job is associated with a text document describing the resume or job requirements.
Besides, they are also associated with a set of (directed) interaction records (\eg job application or resume review) in the job hunting platform.
These interaction sets are formally denoted by $\mathcal{A}_{c_i} =\{c_i \rightarrow j' | c_i \in \mathcal{C}, j' \in \mathcal{J}\}$ and $\mathcal{A}_{j_k} =\{ j_k \rightarrow c' |j_k \in \mathcal{J}, c' \in \mathcal{C}\}$, described as  directed interactions or links requested by candidate $c_i$ or employer $j_k$ (\aka job). 
Here, we use $i$ and $k$ as the indices for candidate and job, respectively. 

Through these interaction behaviors, employers  and candidates might reach agreements on some job positions, which form a matching set 
$\mathcal{M}=\{(c_i, j_k)| c_k \in \mathcal{C}, j_k \in \mathcal{J}\}$, denoting all matching pairs in the platform.
It should be noted that the matching pairs are removed from the interaction sets of both sides.
Actually, a matching record $(c_i, j_k)$ can also be considered as two directed interaction records, namely $c_i\rightarrow j_k$ and $j_k \rightarrow c_i$. 
Based on the interactions and matching results, the task of person-job fit is to learn a matching function $y=f(c_i, j_k)$ that predicts the matching results or score $y$ between the job-person pair~\cite{shen2018joint}, which can be equally cast as a ranking task \emph{w.r.t.} to a job or candidate~\cite{le2019towards}.

In job recruitment, offering and accepting a position is indeed a two-way process~\cite{mcnamara1990job}, where both the candidate and employer can express their intention to each other, called \emph{selection preference}. In prior person-job fit studies~\cite{zhu2018person}, they mainly consider the overall matching results (either success or failure), while the two-way selection preference is seldom considered in the matching function. 
For this purpose, we explicitly model 
candidate- and employer-specific selection preference, denoted by $r_{i \rightarrow k}$ and $s_{k \rightarrow i}$, indicating the intention degree of candidate $c_i$ on job $j_k$ and vice versa. In this way, we aim to incorporate \emph{dual-perspective selection} (candidate's perspective and employer's perspective) for improving the performance of person-job fit. \textcolor{black}{We present the used notations throughout the paper in Table~\ref{tab:notations}.}

\begin{table*}[!t]
\caption{Notations and explanations.}\label{tab:notations}
\begin{tabular}{c|l}
  \hline
{\bf Notation} & \multicolumn{1}{c}{\bf Explanation} \\
\hline
\hline
$c_i, c', c_1, c_2$ & candidates \\
$j_k, j', j_1, j_2$ & jobs \\
$\mathcal{C}$ & the set of candidates \\
$\mathcal{J}$ & the set of jobs \\
$\mathcal{A}_{c_i}$ &  the interaction set that the candidate $c_i$ actively interact \\
$\mathcal{A}_{j_k}$ & the interaction set that the job $j_k$ actively interact \\
$c_i\rightarrow j_k, j_k \rightarrow c_i$ & candidate $c_i$ interact with job $j_k$ actively and passively \\
$r_{i \rightarrow k}$ & candidate-specific selection preference score from candidate $c_i$ to job $j_k$\\
$s_{k \rightarrow i}$ & employer-specific selection preference score from job $j_k$ to candidate $c_i$\\
${d_E}$ & the dimension of the preference embedding\\
${d_O}$ & the output dimension of BERT for document encoding \\
${d_T}$ & the dimension of the document representation \\
$\bm{c}^{a}_i, \bm{c}^{p}_i \in \mathbb{R}^{d_E}$ & active and passive representations of a candidate $c_i$ \\
$\bm{j}^{a}_k, \bm{j}^{p}_k \in \mathbb{R}^{d_E}$ & active and passive representations of a job $j_k$ \\
$\bm{t}_n \in \mathbb{R}^{d_O}$ & the document representation of a resume or a job description\\
$\bm{z}^{(0)}_n \in \mathbb{R}^d$ & the initial representation for node $n$\\
$\bm{W} \in \mathbb{R}^{d_T \times d_O}$ & a learnable transformation matrix for node representation initialization \\ 
$\bm{z}_n \in \mathbb{R}^d$ & the final representation encoded via graph neural network for node $n$ \\
$\omega$ & a hyper-parameter to balance the two types of propagation in graph convolution \\
$\mathcal{M}_{n}$ & the matching set related to node $n$ \\
$\mathcal{N}_n$ & the neighbors of node $n$ \\
$\tau$ & the temperature hyper-parameter in dual-perspective contrastive learning optimization function\\
$\lambda$ & a hyper-parameter to control the strength of the contrastive loss at the training stage\\

 \hline
 \end{tabular}
\end{table*}

\subsection{Dual-Perspective Graph Representation Learning for Directed Selection Intention}

As described before, the person-job fit requires considering the two-way selection process, and thus we propose a  dual-perspective graph convolution network for modeling the directed interactions between jobs and candidates. We first construct the dual-perspective interaction graph and then design a specific graph convolution layer to learn high-level intention representations for both candidates and jobs.

\subsubsection{Constructing the Dual-Perspective Interaction Graph}\label{sec:dpg_construct}

To model the directed behavior for person-job fit, we propose to construct the dual-perspective interaction graph. 

Specifically, given a candidate $c$, we denote its active and passive representations by $\bm{c}^{a}$ (describing the active selection preference) and $\bm{c}^{p}$ (describing the passive selection preference), respectively. Similarly, let $\bm{j}^{a}$ and $\bm{j}^{p}$ denote the active and passive representations of job $j$.
Each representation corresponds to one node in the graph, that is, each candidate (job) is represented with two nodes. These two kinds of nodes aim to capture the node characteristics to actively select others (active perspective) or be passively selected by others (passive perspective). Intuitively, the final match can only be achieved when the intention scores from both active and passive perspectives are strong. 

The edges are constructed by the different interactions,  as well as the association between two nodes of the same candidate (job).
Since such a characterization is symmetric, we can model these directed interactions by modeling the following three cases. Examples of these three different interactions and the corresponding edges are illustrated in Fig~\ref{fig:model}.


$\bullet$ \emph{A candidate applied for a job but was not accepted.} 
It means that the job meets the expectations of the candidate, but the candidate doesn't meet the expectations of the employer. In this case, such an interaction reflects candidate's active preference. Thus, we add an edge between nodes whose representations are $\bm{c}^{a}$ and $\bm{j}^{p}$.

$\bullet$ \emph{An employer reached out to a candidate but was refused.} 
Similar to the first case,
an edge between nodes with representation $\bm{j}^{a}$ and $\bm{c}^{p}$ is established.

$\bullet$ \emph{The two sides reached an agreement for offline interviews.} 
It indicates that the two sides meet the expectation of each other. Under this circumstance, we connect the corresponding active and passive nodes of the involved candidate and job ($\bm{c}^{a}$ and $\bm{j}^{p}$, $\bm{j}^{a}$ and $\bm{c}^{p}$), forming two edges for matching. 

\subsubsection{Initializing the Node Representations}\label{sec:init}

To initialize the node representations, for each node $n$ in the dual-perspective interaction graph, we first apply a look-up table operation for obtaining a preference embedding $\bm{e}_n \in \mathbb{R}^{d_E}$, where $d_E$ is the dimension of the preference embeddings and this embeddings are generated randomly at first.

Besides, each node is associated with a descriptive text (\ie job description or candidate resume). This text information provides important evidence to understand and model the node characteristics. Therefore, we further utilize the BERT model~\cite{devlin2019bert} to encode the 
corresponding text for deriving node representations. Specially, we keep the original order of the text and insert a special token \texttt{[CLS]} before the text. Then the concatenated sequence is fed to the BERT model and the document representation $\bm{t}_n \in \mathbb{R}^{d_O}$ can be obtained, where $d_O$ is the output dimension of BERT.

\ignore{One candidate' resume and her preferences are all key features to indicate whether she will match a job well and vice versa.
We fuse the text representations with users' preference representations together as nodes' initialize representations of the bilateral interaction graph.
For a given node $n$, we obtain its preference embedding $\bm{e}_n \in \mathbb{R}^{d_e}$ via a look-up table, where $d_e$ is the dimension of preference embeddings.
Then we 
adopt BERT~\cite{devlin2019bert} as text encoder to obtain its text representation.
Formally, given a resume (or job description), we firstly concatenate a special symbol \texttt{[CLS]} with the document in order and derive the input sequence for BERT. Then the concatenated sequence is fed to the BERT model and obtain the document representation $\bm{t}_i \in \mathbb{R}^{d_o}$, where $d_o$ is the output dimension of BERT.
}

Finally, the two representations mentioned above are fused as the initial representation for node $n$:
\begin{align}
    \bm{z}_n^{(0)} = [\bm{e}_n; \bm{W}\cdot \bm{t}_{n}] \in \mathbb{R}^{d},\label{equ:emb}
\end{align}
where $\bm{W} \in \mathbb{R}^{d_T \times d_O}$ is a learnable transformation matrix, $[;]$ denotes concatenation operation and $d = d_E + d_T$ is dimension of the initialized node representations.



\subsubsection{Propagating the Hybrid Preference}

Since we have characterized the interactions in the form of the interaction graph, we adopt the graph convolution networks~(GCN) to learn the node representations. Different from prior GCN studies~\cite{he2020lightgcn}, we have two kinds of different edges for each involved candidate and job. Therefore, we propose a hybrid preference propagation algorithm for learning the node representations. 
It should be noted that the differences are inherent in edge rather than node type, so the preference propagation of all nodes can be defined uniformly.

At the $l$-th layer of graph convolution, for each node $n$
(either job or candidate, either active or passive),
we consider the preference propagation from two different sets of interactions: the matching set related to node $n$ denoted by $\mathcal{M}_n$ and the interaction set related to node $n$ denoted by $\mathcal{A}_n$. Formally, we adopt a lightweight propagation mechanism~\cite{he2020lightgcn} for updating the node representations: 
\begin{align}
\bm{z}_n^{(l)}=\sum\limits_{u \in \mathcal{M}_n} \frac{1}{\sqrt{|\mathcal{N}_n||\mathcal{N}_u|}} \bm{z}_u^{(l - 1)} + \omega \sum\limits_{v \in \mathcal{A}_n} \frac{1}{\sqrt{|\mathcal{N}_n||\mathcal{N}_v|}} \bm{z}_v^{(l - 1)},\label{eq:gnn_prop}
\end{align}
where $\bm{z}_u$ and $\bm{z}_v$ denote the neighbors that have matching interaction and unidirectional interaction, respectively, and
$\mathcal{N}_n$, $\mathcal{N}_{u}$ and $\mathcal{N}_{v}$ are neighbors of node $n$, node $u$ and node $v$.
Here,  we incorporate a specific hyper-parameter $\omega$ to balance the two types of propagation, since these two kinds of interactions convey different levels of preference
when learning node representations.


We average the representations from $(L+1)$ layers as the final representation for each node $n$ as follows:
\begin{align}
\bm{z}_n = \sum_{l = 0}^{L} \frac{1}{L+1}\bm{z}_n^{(l)},
\end{align}
where $L$ denotes the number of graph convolution layers.

\subsection{Prediction with Two-Way Intentions}
Above, we have constructed the interaction graph and introduced the propagation mechanism for updating node representations. Next, we discuss how to learn the model parameters according to the person-job matching records.

After learning node representations, we can compute  the two-way selection preference  $r_{i \rightarrow k}$ (the intention that a candidate $c_i$ selects a job $j_k$) and $s_{k \rightarrow i}$ (the intention that a job $j_k$ selects a candidate $c_i$) for modeling the intention from dual perspectives. Formally, given a candidate $c_i$ and a job $j_k$, we use the inner products to compute these two kinds of intention scores:
\begin{equation}
\begin{aligned}
    r_{i \rightarrow k} = \bm{c}^a_i \cdot \bm{j}^p_k,  \\
    s_{k \rightarrow i} = \bm{j}^a_k \cdot \bm{c}^p_i.\label{eq:constraints}
\end{aligned}
\end{equation}

Finally, we integrate the two intention degrees to predict the matching score: 
\begin{align}
    \hat{y}=\frac{1}{2} r_{i \rightarrow k} + \frac{1}{2}s_{k \rightarrow i}.
\end{align}

Different from prior person-job fit studies (either modeling the overall or unidirectional fit), our approach explicitly models the two-way selection preference, which is expected to better capture the actual recruitment process. 




\subsection{Self-Supervised Enhanced Dual-Perspective Ranking Optimization}

For person-job fit, it is more difficult to directly predict an absolute matching score, which is likely to result in over-fitting and model bias~\cite{le2019towards}. Therefore, we adopt a ranking-based approach to optimize the entire model. 


\subsubsection{Quadruple-based Loss Function}

Unlike previous studies, which either optimize a cross-entropy loss~\cite{qin2018enhancing} or a pairwise comparison loss~\cite{le2019towards}, we introduce a novel quadruple-based loss function.
\textcolor{black}{
In the recruitment scenario, a successful matching often means that the job position should rank highly from the perspective of the candidate, and vice versa.
Both perspectives are equally important, and both must be satisfied at the same time.
}

Given a matched record $\langle c_i, j_k \rangle$, we construct a quadruple 
$\langle c_i, j_k, c_{i'}, j_{k'} \rangle$ by sampling  
a non-matched candidate and a non-matched job, \ie constructing the negative pairs $\langle c_i, j_{k'} \rangle$ and $\langle c_{i'}, j_{k} \rangle$ for optimization. 
Intuitively, the matching scores should hold the following conditions: 
\begin{equation}\label{eq-comparison}
\begin{aligned}
    f(c_i, j_k) > f(c_i, j_{k'}), \\
   f(c_i, j_k) > f(c_{i'}, j_{k}).
\end{aligned}
\end{equation}
For simplicity, we abbreviate $\langle c_i, j_k, c_{i'}, j_{k'} \rangle$ by  $\langle i, k, i', k' \rangle$. Then, we extend the widely used BPR loss~\cite{rendle2009bprmf} by modeling the quadruple-based loss as follows:
\begin{align}
    \mathcal{L}_{main} = -\frac{1}{|{\mathcal{D}}|}\sum_{(i, k, i^{'}, k^{'}) \in {\mathcal{D}}}\log \bigg(\sigma\big(\hat{y}_{i,k} -\frac{1}{2} \hat{y}_{i,k^{'}} -\frac{1}{2}  \hat{y}_{i^{'},k}\big) \bigg) \label{eq:l_main},
\end{align}
where  $\mathcal{D} = \{(i,k,i^{'},k^{'})|(i,k)\in \mathcal{M}, (i,k^{'}) \in \mathcal{M}^{-},$ $(i^{'}, k) \in \mathcal{M}^- \}$ denotes the training data,
$\mathcal{M}$ and $\mathcal{M}^-$ denote the matched and unmatched sets, respectively. Here, we average the scores of $\hat{y}_{i,k^{'}}$ and $\hat{y}_{i^{'},k}$ to constrast with the matched score $\hat{y}_{i,k}$ via the sigmoid function $\sigma(\cdot)$. As we can see, such a loss tries to optimize the partial orders in Eqn.~\eqref{eq-comparison}.

\subsubsection{Dual-Perspective Contrastive Learning}

Besides the above optimization, we further consider another natural constraint for optimizing the model parameters. Since we set up two kinds of representations for each candidate (or job), the two representations should be similar to each other. 
Inspired by the above insight, we design 
a dual-perspective contrastive learning optimization function. Specifically, 
we treated the \emph{active representation} and the \emph{passive representation} of the same candidate (or job) as the positive pairs (\emph{i.e.}, $\{(\bm{c}_i^a, \bm{c}_i^p)| c_i \in \mathcal{C}\}$), 
and the representations
between candidates (or jobs)
as the negative pairs (\emph{i.e.}, $\{(\bm{c}_i^a, \bm{c}_{i^{'}}^p)| c_i, c_{i^{'}} \in \mathcal{C}\}$).
The auxiliary supervision of positive pairs promotes the consistency of representations from different perspectives for the same candidate (or job), while the supervision of negative pairs tries to enlarge the divergence among different subjects.
Formally, we adopt the InfoNCE~\cite{oord2018representation} loss to maximize the agreement of positive pairs and minimize that of negative pairs:
\begin{align}
\mathcal{L}^{C}_{ssl} = - \sum_{c_i \in \mathcal{C}} \log{ \frac{\exp({(\bm{c}_i^a \cdot \bm{c}_i^p) / \tau})}{\sum\limits_{c_{i^{'}} \in \mathcal{C}}(\exp({(\bm{c}_i^a\cdot \bm{c}_{i^{'}}^p) / \tau}) + \exp({(\bm{c}_{i^{'}}^a\cdot \bm{c}_i^p)/ \tau}))}},\label{eq:ssl}
\end{align}
where $\tau$ is the temperature hyper-parameter.
Similarly, we can obtain the contrastive loss on the job side $\mathcal{L}^{J}_{ssl}$. Combining these two losses, the final objective function of the self-supervised task is
given as:
\begin{align}
    \mathcal{L}_{ssl} = \mathcal{L}^{C}_{ssl} + \mathcal{L}^{J}_{ssl}. \label{eq:l_cl}
\end{align}

\subsubsection{Training and Complexity Analysis}
At the training stage, we leverage a multi-task training strategy to jointly optimize the proposed quadruple-based ranking loss (Eqn.~\eqref{eq:l_main}) and dual-perspective contrastive loss (Eqn.~\eqref{eq:l_cl}):
\begin{align}
    \mathcal{L} = \mathcal{L}_{main} + \lambda \mathcal{L}_{ssl},
    \label{eq:l_joint}
\end{align}
where $\lambda$ is a hyper-parameter to control the strength of the individual-level contrastive learning task.

For training the proposed model, the  major time consuming part lies in the propagation via GCN (Eqn.~\eqref{eq:gnn_prop}) and calculation of $\mathcal{L}_{ssl}$ (Eqn.~\eqref{eq:l_cl}). 
For propagation of GNN, it requires a time of $O(|\mathcal{E}| \cdot L \cdot d)$, where $\mathcal{E} = \mathcal{D} \cup \mathcal{A}^j \cup \mathcal{A}^c$ denotes the set of all edges in the constructed dual-perspective interaction graph, $L$ is the number of GNN layers and $d$ is the dimension of node representations in GNN.
For the dual-perspective contrastive loss, we usually sample $S$ negative candidates (or jobs) for each user in practice. Then the time complexity can be roughly estimated as $O(|\mathcal{V}|\cdot S \cdot d)$, where $\mathcal{V} = \mathcal{C} \cup \mathcal{J}$ denotes the set of all nodes es in the interaction graph.
Note that once users submit their resumes or job description, we can obtain the corresponding BERT representations offline, and thus we ignore the corresponding cost here. 
Overall, the time complexity for training an epoch is $O(|\mathcal{E}| \cdot L \cdot d + |\mathcal{V}|\cdot S \cdot d)$. 
The full algorithm is described in Algorithm~1.

\begin{imageonly}
\begin{algorithm2e}[!t]
  \caption{The training algorithm of \textbf{DPGNN}.}
  \label{algo:dpgnn}
  \DontPrintSemicolon\SetNoFillComment
  \SetKwInOut{Input}{input}
\small
  \Input{candidate set $\mathcal{C}$ and job set $\mathcal{J}$, as well as their associated text document and directed interactions, and matching set $\mathcal{M}$ defined in Sec.~\ref{sec:not}.}
  \BlankLine
  Construct the dual-perspective interaction graph with directed behaviors in training set. (Sec.~\ref{sec:dpg_construct})\;
  Initialize preference embedding $\bm{e}_n$ and document representation $\bm{t}_n$ for each node $n$. (Sec.~\ref{sec:init})\;
  \ForEach {mini-batch $\mathcal{B}$}{
    Update node representations $\bm{z}_n^{(0)}$ for each node $n$ with Eqn.~\eqref{equ:emb}\;
    Obtain the final node representations $\bm{z}_n$ via propagating the hybrid preference with Eqn.~\eqref{eq:gnn_prop}\;
    $\mathcal{L}_{main} \gets 0$ \;
    \ForEach {candidate-job pair $\langle c_i, j_k \rangle \in \mathcal{B}$}{
        Sample a non-matched candidate $c_{i'}$ (for $j_k$) and a non-matched job $j_{k'}$ (for $c_i$) \;
        $\hat{y}_{i,k} \gets \frac{1}{2} \bm{c}^{a}_i \cdot \bm{j}^{p}_k + \frac{1}{2} \bm{j}^a_k \cdot \bm{c}^p_i$ (similar for $\hat{y}_{i, k'}$ and $\hat{y}_{i', k}$) \;
        $\mathcal{L}_{main} \gets \mathcal{L}_{main} - \log(\sigma(\hat{y}_{i,k} -\frac{1}{2} \hat{y}_{i,k'} -\frac{1}{2} \hat{y}_{i',k}))$\;
    }
    Calculate $\mathcal{L}_{ssl}^C$ with candidates in this mini-batch with Eqn.~\eqref{eq:ssl} (similar for $\mathcal{L}_{ssl}^J$)\;
    $\mathcal{L} = \mathcal{L}_{main} + \lambda \left(\mathcal{L}^{C}_{ssl} + \mathcal{L}^{J}_{ssl}\right)$\;
    Backpropagate $\mathcal{L}$ and update model parameters\;
  }
\end{algorithm2e}\end{imageonly}

\subsubsection{Discussion}
We conduct a discussion here with comparisons to existing methods for a better understanding of the proposed optimization techniques.
For the proposed quadruple-based ranking loss,
the negative examples are sampled from both perspectives to meet the ordering constraints in Eqn.~\eqref{eq:constraints}.
Optimizing the quadruple-based object makes the dual-perspective ranking stable.
Besides, inspired by recent advances in self-supervised recommendation~\cite{yu2022self}, we propose the dual-perspective contrastive loss to enhance the whole optimization.
Unlike previous self-supervised signals based on attribute correlation~\cite{zhou2020s3rec} and data augmentation~\cite{wu2021sgl}, we consider the natural correlations between the preference representations from two perspectives of the same user (\eg $\bm{c}_i^a$ and $\bm{c}_i^p$ of candidate $c_i$ in Eqn.~\eqref{eq:ssl}) as self-supervised supervision, which is both efficient and effective.

\section{Experiment}

\textcolor{black}{
In this section, we would like to evaluate the performance of the proposed DPGNN. 
Extensive experiments conducted on three real-world datasets aim to answer the following research questions:
}
\begin{itemize}
    \item \textbf{RQ1}: Does our model outperform the content-based, collaborative filtering-based, and hybrid state-of-the-art methods on the person-job fit task with the real-world datasets?
    \item \textbf{RQ2}: How do the key components (\eg interaction graph, quadruple-based loss, and dual-perspective contrastive loss) of our model benefit the prediction?
    \item \textbf{RQ3}: How does our model perform with users who have varied levels of interaction sparsity?
    \item \textbf{RQ4}: How to select hyper-parameters for better performance of the proposed method?
\end{itemize}
\textcolor{black}{
In what follows, we first set up the experiments, and then present and analyze the results. Finally, we present how the proposed method performs in a specified case.
}

\begin{table}[]
\caption{Statistics of the experiment dataset.}
\label{tab:data_statistics}
\resizebox{1.0\columnwidth}{!}
{
\begin{tabular}{@{}crrrrr@{}}
\toprule
Statistics & \# Candidates & \# Jobs  & \# C Interaction & \# J Interaction & \# Match\\ \midrule
Tech       & 56,634       & 48,090  & 3,749,807              & 907,087              & 925,193    \\
Sales       & 15,854       & 12,772  & 213,860                & 2,077,560            & 145,066    \\
Design     & 12,290       & 9,143   & 1,200,590              & 76,287               & 166,270    \\ \bottomrule
\end{tabular}
}
\end{table}

\begin{table*}[]
\centering
\caption{Performance comparison of all methods. The best performance and the second best performance methods are denoted in bold and underlined fonts respectively. ``${*}$'' indicates that the improvements are significant at the level of 0.01 with paired $t$-test.
``$-$'' denotes that the corresponding experiment is not finished after $24$ hours' training.}
\label{tab:overall_comparison}
\resizebox{2.0\columnwidth}{!}
{
\begin{tabular}{@{}cccccccccc@{}}
\toprule
\multirow{2}{*}{Dataset}  & Direction    & \multicolumn{4}{c}{Candidates}                                            & \multicolumn{4}{c}{Jobs}                                                  \\ \cmidrule(l){2-10} 
                          & Metric       & Recall@5         & Precision@5            & NDCG@5           & MRR@5            & Recall@5         & Precision@5            & NDCG@5           & MRR@5            \\ \midrule
\multirow{11}{*}{Tech}    & BPRMF        & 0.2544           & 0.0906           & 0.2630           & 0.2692           & 0.2987           & 0.1464           & 0.3514           & 0.3403           \\
                          & NCF          & 0.2496           & 0.0902           & 0.2646           & 0.2707           & 0.3040           & 0.1542           & 0.3727           & 0.3609           \\
                          & LightGCN     & 0.2557           & 0.0925           & 0.2703           & 0.2762           & 0.3076           & 0.1535           & 0.3674           & 0.3545           \\
                          & LFRR         & 0.2366           & 0.0817           & 0.2308           & 0.2376           & 0.2776           & 0.1228           & 0.3009           & 0.2856           \\ \cmidrule(l){2-10} 
                          & PJFNN        & 0.2104           & 0.0770           & 0.2232           & 0.2362           & 0.2733           & 0.1404           & 0.3578           & 0.3439           \\
                          & BPJFNN       & 0.2136           & 0.0727           & 0.2033           & 0.2167           & 0.2620           & 0.1264           & 0.3146           & 0.3008           \\
                          & APJFNN       & -                & -                & -                & -                & -                & -                & -                & -                \\ \cmidrule(l){2-10} 
                          & $\text{LGCN}_{\text{BERT}}$ & \underline{0.2685}     & \underline{0.1003}     & \underline{0.2951}     & \underline{0.2982}     & 0.3187           & \underline{0.1713}     & \underline{0.4085}     & \underline{0.3914}     \\
                          & IPJF         & 0.2591           & 0.0941           & 0.2654           & 0.2699           & 0.2979           & 0.1606           & 0.3818           & 0.3644           \\
                          & PJFFF        & 0.2556           & 0.0985           & 0.2785           & 0.2791           & \underline{0.3195}     & 0.1691           & 0.4049           & 0.3870           \\
                          & DPGNN        & \textbf{0.2941*} & \textbf{0.1076*} & \textbf{0.3119*} & \textbf{0.3100*} & \textbf{0.3430*} & \textbf{0.1858*} & \textbf{0.4410*} & \textbf{0.4167*} \\ \midrule
\multirow{11}{*}{Sales}   & BPRMF        & 0.2041           & 0.0643           & 0.1966           & 0.2156           & 0.2462           & 0.0843           & 0.2320           & 0.2431           \\
                          & NCF          & 0.2031           & 0.066            & 0.1950           & 0.2138           & 0.2332           & 0.0796           & 0.2233           & 0.2395           \\
                          & LightGCN     & 0.2084           & 0.0657           & 0.1966           & 0.2152           & 0.2575           & 0.0903           & 0.2473           & 0.2538           \\
                          & LFRR         & 0.2244                 &   0.0703               &    0.2064              &  0.2213                &  0.2397                & 0.0839                 &  0.2312                &   0.2418               \\ \cmidrule(l){2-10} 
                          & PJFNN        & 0.1385           & 0.0422           & 0.1180           & 0.1500           & 0.1550           & 0.0484           & 0.1396           & 0.1689           \\
                          & BPJFNN       & 0.1641           & 0.0521           & 0.1522           & 0.1777           & 0.2015           & 0.0676           & 0.1879           & 0.2065           \\
                          & APJFNN       & 0.1756           & 0.0588           & 0.1701           & 0.1908           & 0.2187           & 0.0722           & 0.2013           & 0.2198           \\ \cmidrule(l){2-10} 
                          & $\text{LGCN}_{\text{BERT}}$ & \underline{0.2272}     & \underline{0.0731}     & \underline{0.2184}     & \underline{0.2310}     & \underline{0.2518}     & \underline{0.0891}     & \underline{0.2404}     & \underline{0.2509}     \\
                          & IPJF         & 0.1393           & 0.0436           & 0.1244           & 0.1571           & 0.1977           & 0.0672           & 0.1955           & 0.2140           \\
                          & PJFFF        & 0.1973           & 0.0644           & 0.1869           & 0.2059           & 0.2478           & 0.0858           & 0.2363           & 0.2445           \\
                          & DPGNN        & \textbf{0.2330*} & \textbf{0.0765*} & \textbf{0.2232*} & \textbf{0.2373*} & \textbf{0.2578*} & \textbf{0.0916*} & \textbf{0.2591*} & \textbf{0.2664*} \\ \midrule
\multirow{11}{*}{Design} & BPRMF        & 0.2349           & 0.0726           & 0.2252           & 0.2417           & 0.2312           & 0.0919           & 0.2478           & 0.2501           \\
                          & NCF          & 0.2386           & 0.0761           & 0.2366           & 0.2505           & 0.2410           & 0.1014           & 0.2728           & 0.2733           \\
                          & LightGCN     & 0.2451           & 0.0756           & 0.2336           & 0.2457           & 0.2438           & 0.1019           & 0.2707           & 0.2711           \\
                          & LFRR         &  0.2342                &  0.0699                &  0.2105                &  0.2272                &  0.2478                &  0.1042                &  0.2789                &  0.2780                \\ \cmidrule(l){2-10} 
                          & PJFNN        & 0.1592           & 0.0518           & 0.1597           & 0.1886           & 0.1201           & 0.0511           & 0.1445           & 0.1661           \\
                          & BPJFNN       & 0.1725           & 0.0539           & 0.1568           & 0.1826           & 0.1797           & 0.0694           & 0.1874           & 0.2003           \\
                          & APJFNN       & 0.2120           & 0.0624           & 0.1738           & 0.2054           & 0.2022           & 0.0843           & 0.2314           & 0.2413           \\ \cmidrule(l){2-10} 
                          & $\text{LGCN}_{\text{BERT}}$ & \underline{0.2517}     & \underline{0.0800}     & \textbf{0.2498*}  & \textbf{0.2637*}  & 0.2567           & 0.1158           & 0.2998           & 0.2934           \\
                          & IPJF         & 0.2487           & 0.0798           & 0.2445           & 0.2553           & 0.2294           & 0.1159           & 0.3013           & 0.2987           \\
                          & PJFFF        & 0.2400           & 0.0766           & 0.2367           & 0.2484           & \underline{0.2582}     & \underline{0.1193}     & \underline{0.3156}     & \underline{0.3121}     \\
                          & DPGNN        & \textbf{0.2685*} & \textbf{0.0808*} & \underline{0.2478}     & \underline{0.2584}     & \textbf{0.2783*} & \textbf{0.1342*} & \textbf{0.3524*} & \textbf{0.3379*} \\ \bottomrule
\end{tabular}
}
\end{table*}

\subsection{Experiment Setup}
\paratitle{Dataset.}
We evaluate our model on three large real-world datasets provided by a popular online recruiting platform. The datasets were constructed from 106 days' real online logs and contained two kinds of behavior: \emph{Match} and \emph{Interaction}, corresponding to the matching set and interaction set mentioned in Section~\ref{sec:not}.
Besides, each candidate (and job) is associated with a descriptive text (\ie resume or job description).
The overall statistics are shown in Table~\ref{tab:data_statistics}.

\paratitle{Baseline Models.}
We compare our method with the following
baseline models:
\begin{itemize}
\item {\textbf{BPRMF}~\cite{rendle2009bprmf}} learns the latent user/item embeddings via minimizing Bayesian Personalized Ranking (BPR) loss.
\item {\textbf{NCF}~\cite{he2017neural}} leverages an MLP to replace the inner product to model the interaction between user and item.
\item {\textbf{LightGCN}~\cite{he2020lightgcn}} simplifies GCN's design to make it concise for collaborative filtering.
\item {\textbf{$\text{LGCN}_{\text{BERT}}$}}. It has the same structure as LightGCN, except that the initial embedding is replaced by $[\bm{e}_i ; \text{MLP}(\bm{o}_{T})]$, which is the same as Eqn.~\eqref{equ:emb}.
\textcolor{black}{
\item {\textbf{LFRR}~\cite{neve2019latent}} is a reciprocal collaborative filtering method based on latent factors. 
}
\item {\textbf{PJFNN}~\cite{zhu2018person}} is a method based on convolutional neural network (CNN). Resumes and job descriptions are encoded independently by hierarchical CNN, and the matching degree is calculated by cosine similarity.
\item {\textbf{BPJFNN}~\cite{qin2018enhancing}} leverages bidirectional LSTM to derive the representations of resumes and job descriptions.
\item {\textbf{APJFNN}~\cite{qin2018enhancing}} learns a word-level semantic representation for both resumes and job descriptions based on RNN models with attention hierarchically.
\item {\textbf{IPJF}~\cite{le2019towards}} leverages multiple labels to indicate the propensity of candidates and jobs to reach a match.
\item {\textbf{PJFFF}~\cite{jiang2020learning}} fuses the representations for the explicit and implicit intentions, which are processed by the historical applications using LSTM.
\end{itemize}

All baseline models fall into three categories, depending on the data they focus on:
(1) \emph{Collaborative filtering based methods}:
BPRMF,
NCF, 
LightGCN 
and LFRR; (2) \emph{Content-based methods}: 
PJFNN,
BPJFNN,
APJFNN; (3) \emph{Hybrid methods}: $\text{LGCN}_{\text{BERT}}$, IPJF and PJFFF.
Note that we do not include weaker collaborative filtering-based reciprocal methods as baselines such as RCF~\cite{xia2015reciprocal} and RWS~\cite{kleinerman2018optimally}.
Besides, as content-based methods are indeed reciprocal methods, we do not include additional content-based reciprocal baselines.

\begin{table*}[]
\label{tab:ablation_study}
\caption{Effectiveness analysis of key components of DPGNN. R@5, P@5, NDCG@5, MRR are adopted for evaluation.}
\label{tab:ablation_study}
{
\begin{tabular}{ccccccccc}
\toprule
\multirow{2}{*}{Model}                      & \multicolumn{4}{c}{For candidates}  & \multicolumn{4}{c}{For jobs}      \\ \cmidrule(l){2-9} 
                                            & R@5     & P@5     & NDCG@5 & MRR    & R@5    & P@5    & NDCG@5 & MRR    \\ \midrule
DPGNN                                       & \textbf{0.2941}  & \textbf{0.1076}  & \textbf{0.3119} & \textbf{0.3100} & \textbf{0.3430} & \textbf{0.1858} & \textbf{0.4410} & \textbf{0.4167} \\
w/o DPG                                      & 0.2770  & 0.1008  & 0.2899 & 0.2912 & 0.3360 & 0.1803 & 0.4257 & 0.4045 \\
w/o QL                                      & 0.2756  & 0.1010  & 0.2945 & 0.2964 & 0.3294 & 0.1741 & 0.4188 & 0.3994 \\
w/o SSL                                      & 0.2897  & 0.1072  & 0.3125 & 0.3108 & 0.3346 & 0.1802 & 0.4286 & 0.4075 \\
\bottomrule
\end{tabular}
}
\end{table*}

For each compared method, we learn the model parameters on the training set, perform hyper-parameter selection according to the results on the validation set, and finally report the result on the test set.
Besides, for fair comparisons of those baselines that don't utilize unidirectional interactions, we treat the unidirectional interaction records as positive instances while training.
We observe that it generally improves the performance of these methods,
and we report these results in subsequent experiments.

\paratitle{Evaluation and Implementation Details.}
To evaluate the performance of top-$k$ recommendation~\cite{zhao2021revisiting}, we employ four widely used metrics, Recall (R@$k$), Precision (P@$k$), Normalized Discounted Cumulative Gain (NDCG@$k$) and Mean Reciprocal Rank (MRR). We set $k$ to 5 empirically.
In particular, the evaluation method is different from general recommendation tasks.
We leverage these metrics to do the evaluation for two person-job fit ranking tasks simultaneously, \emph{i.e.,} ranking jobs for candidates and ranking candidates for employers, which is more in line with the real online recruitment scenarios.
Concretely,
for each positive instance $(c_i, j_k)$, we randomly sample $20$ jobs for candidate $c_i$ and $20$ candidates for job $j_k$ as negative instances.
We evaluate the ranking list of the overall matching degree and report the average metric scores over all the candidates and jobs respectively.

The baseline models are implemented with a popular open-source recommendation library \textsc{RecBole}~\cite{zhao2021recbole, zhao2022recbole}.
The interaction records are sorted and split into training, validation and test sets by timestamp. Specifically, the records of the last 11 days and the previous 11 days are used as test set and validation set respectively, and the rest 84 days are used for training.
We optimize all the methods with Adam optimizer and carefully search the hyper-parameters of all the baselines.
Parameters of the text encoders are initialized by \texttt{bert-base-uncased}\footnote{https://github.com/huggingface/transformers}. 
The dimensions of text embeddings and preference embeddings are set to 32 and 128, respectively. 
 The learning rate is tuned in \{0.01, 0.001, 0.0001, 0.00001\}.
Early stopping is used with patience of 10 epochs.

\begin{figure}[t!]
	\centering
	\includegraphics[width=0.4\textwidth]{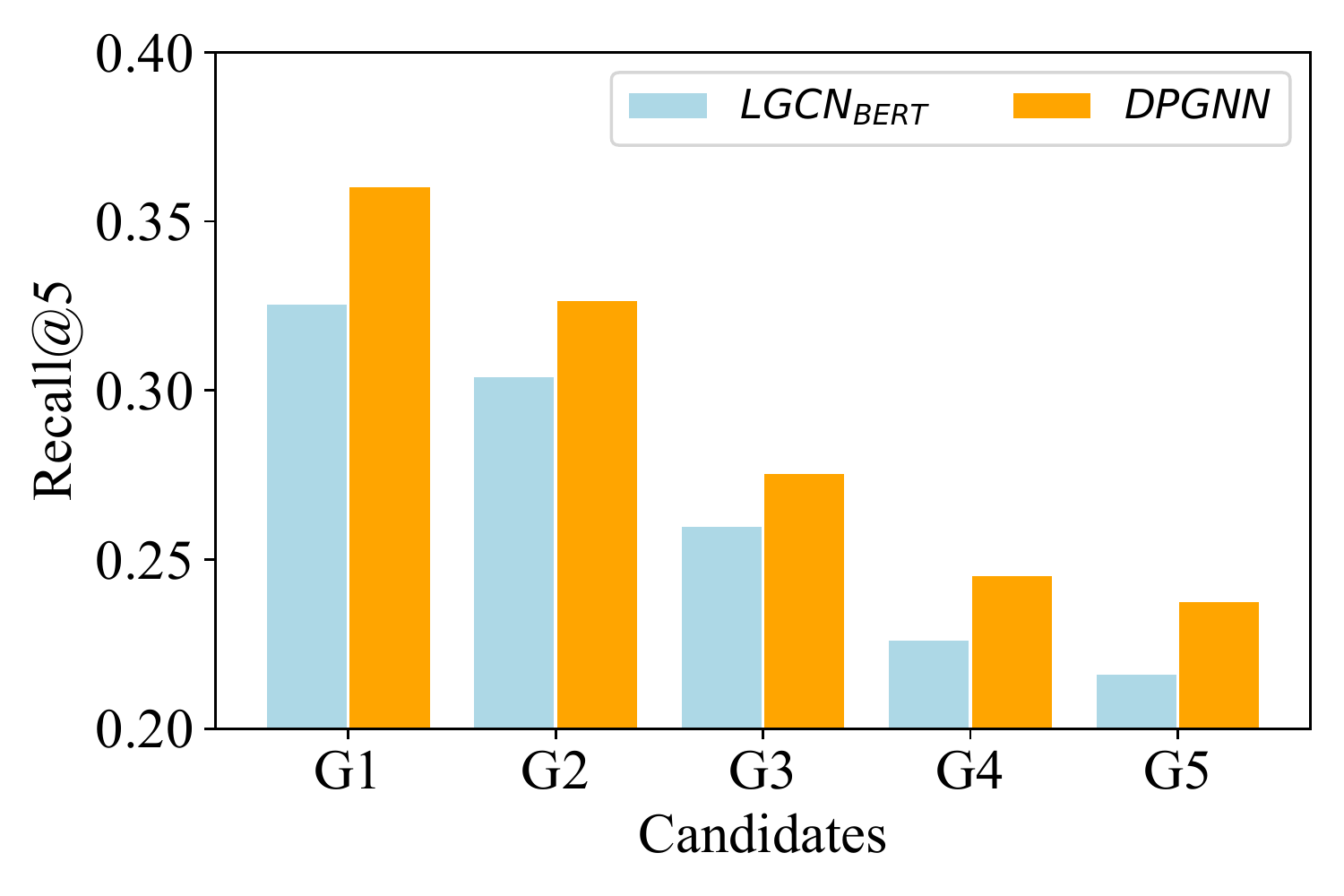}
	\includegraphics[width=0.4\textwidth]{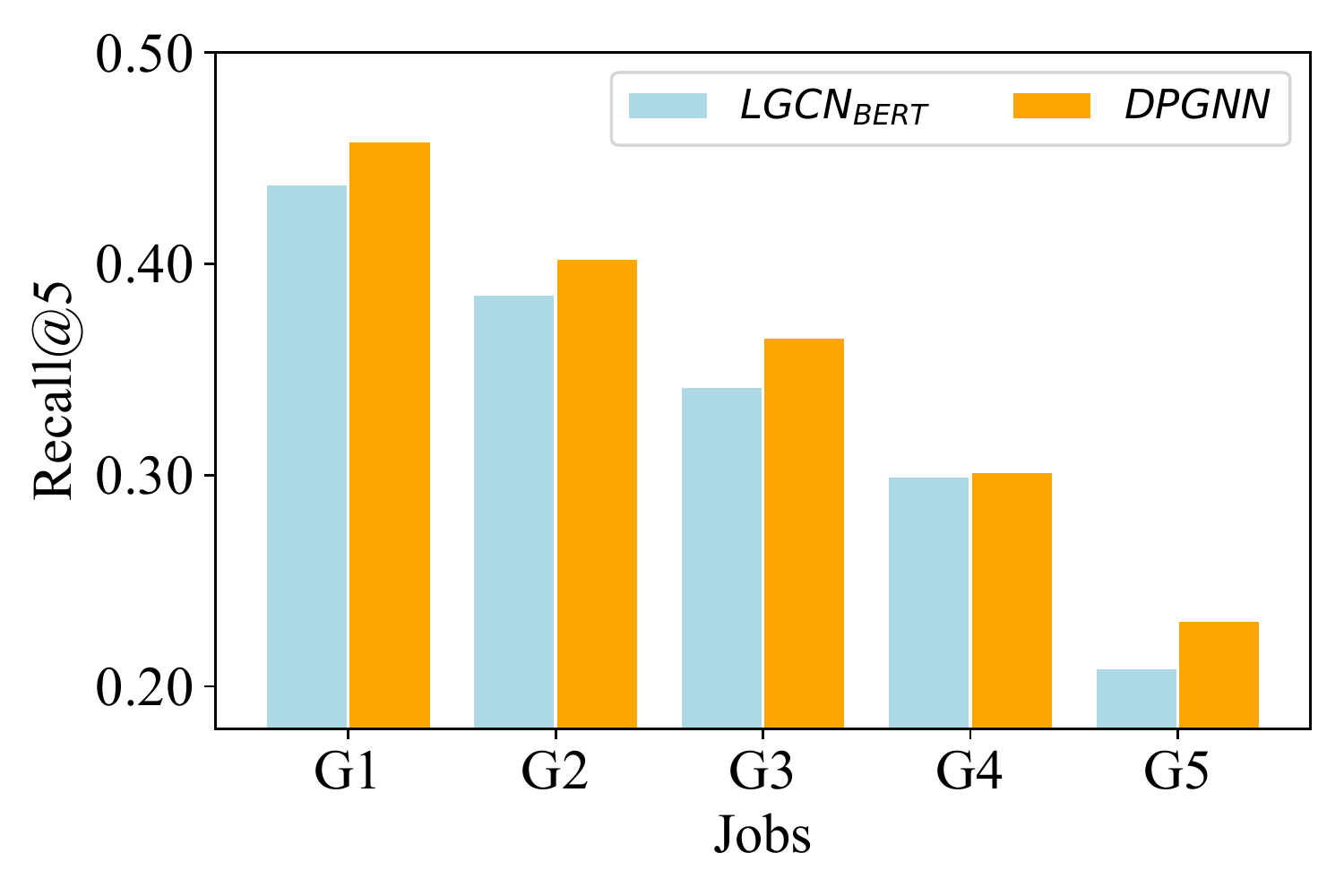}
	\caption{\textcolor{black}{Performance analysis for different sparsity level candidates and jobs,
	G$1$ denotes the group of candidates or jobs with the lowest average number of interactions.}}
	\label{fig:sparsity}
\end{figure}

\begin{figure*}[t!]
	\centering
	\includegraphics[width=0.28\textwidth]{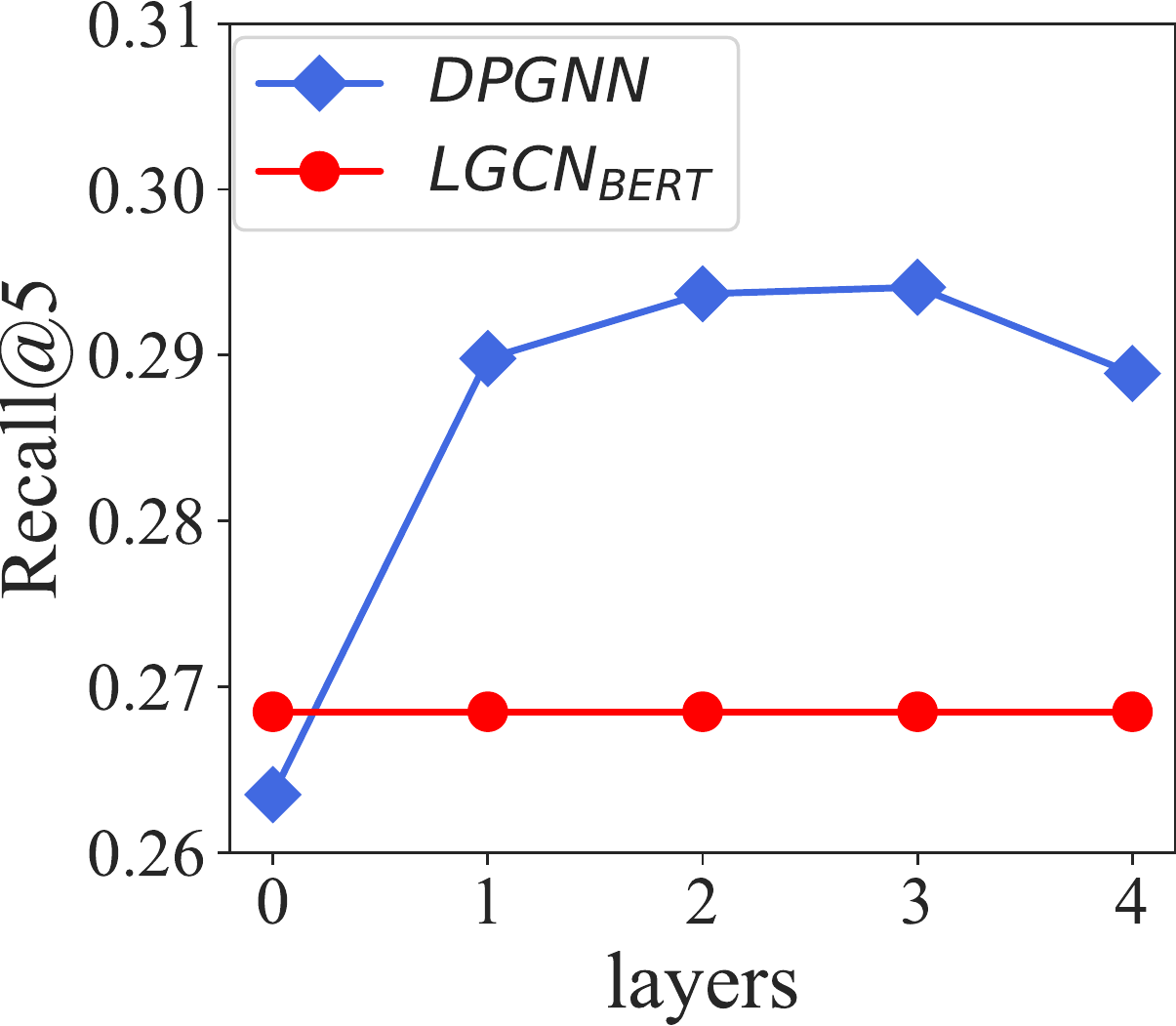}
	\includegraphics[width=0.27\textwidth]{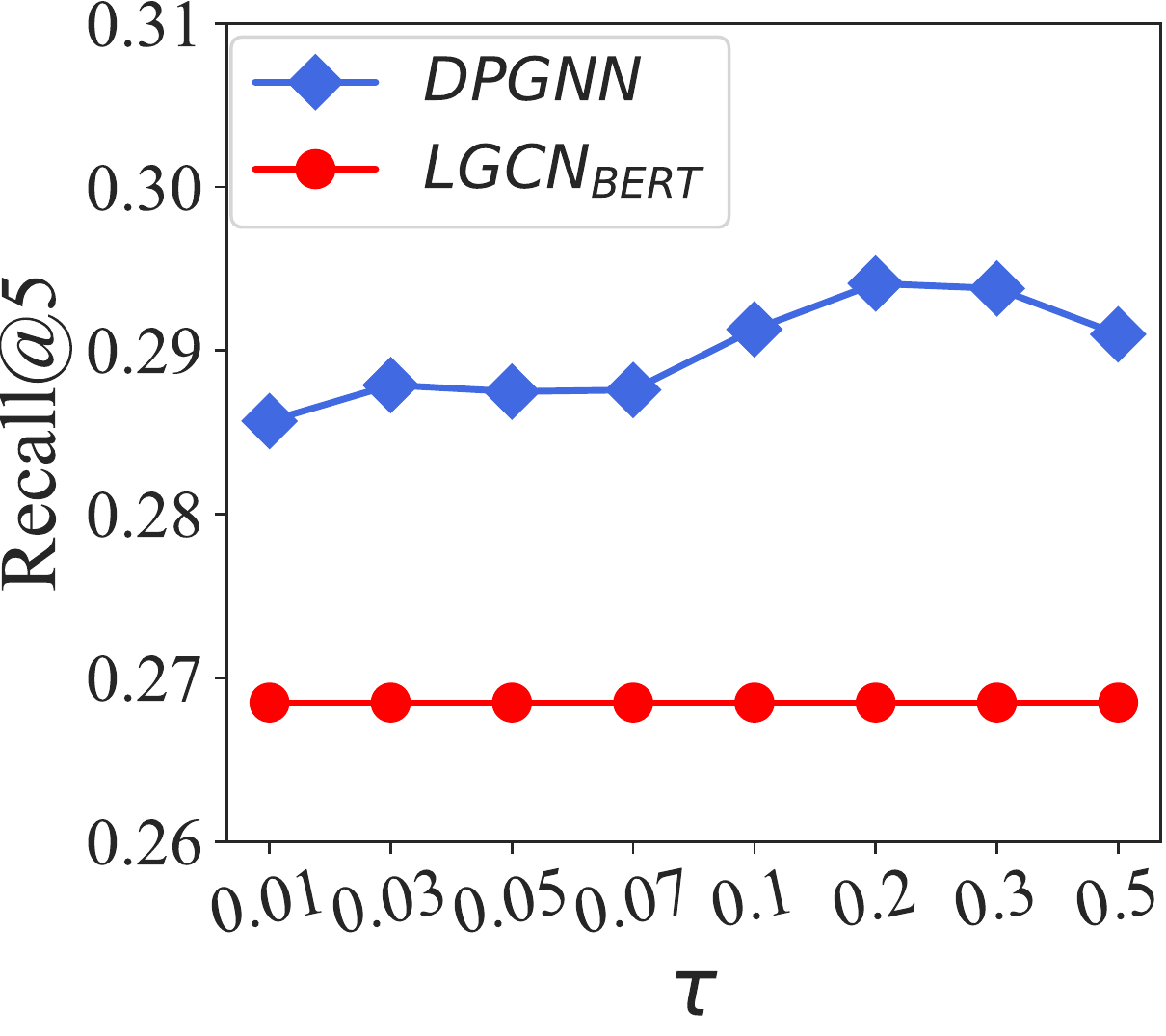}
	\includegraphics[width=0.27\textwidth]{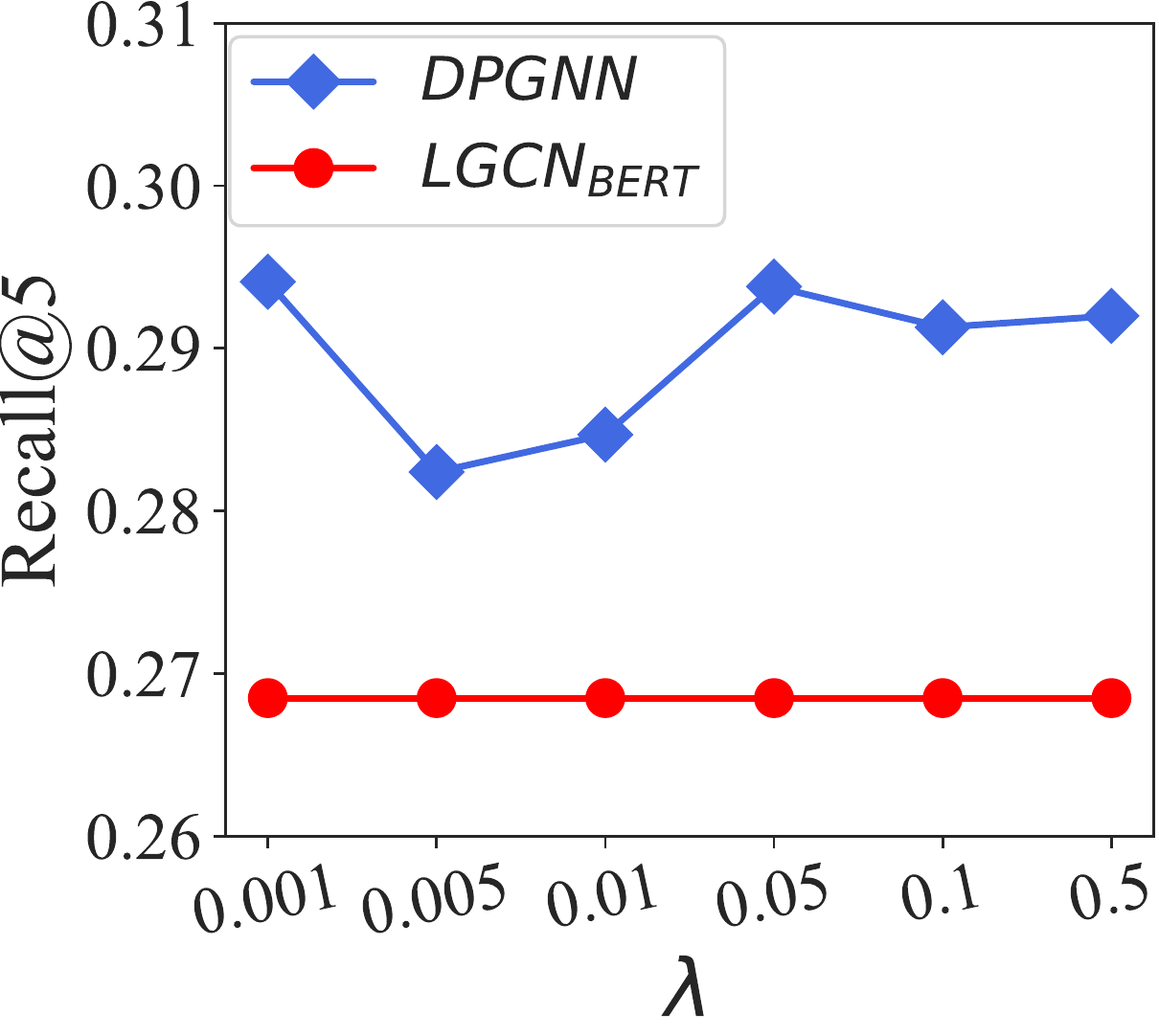}
	\caption{\textcolor{black}{Performance comparison w.r.t. different layers, $\tau$ and $\lambda$.}}
	\label{fig:param_tuning}
\end{figure*}

\subsection{Experiment Result}
\paratitle{Overall Performance (RQ1)}

Table~\ref{tab:overall_comparison} presents the comparison between our model and all the baselines.
For the four collaborative filtering baselines, LightGCN achieves the best performance, but the improvement is not significant compared with BPRMF, NCF and LFRR. 
As for the three content-based baselines, BPJFNN, PJFNN and APJFNN, which highly rely on the text content, they do not perform well in most cases.
The possible reason is that they require resumes and job postings to be structured and unabridged,
while in our scenario, users in the platform have different text organization habits.
IPJF doesn't performs well in Sales, due to the high imbalance in the number of different types of interactions.
The performances of PJFFF are better in most cases, as PJFFF integrates historically interacted resumes or job descriptions.
Besides, documents in Tech usually contain more specific skill requirements than others, making the text-based models relatively efficient in Tech.
Finally, as we can see, $\text{LGCN}_{\text{BERT}}$ leverages both interactions and text. It mostly performs the best across baselines, indicating that utilizing both text descriptions and interactions is important.

As a comparison, our method achieves the best performance on most metrics across three datasets. Specifically, our model can on average improve the best baseline by 7.12\%, 4.81\% and 7.73\% relatively on Tech, Sales and Design, respectively.
Different from baselines, our approach models the two-way selection preferences of both candidates and jobs. Thus, it is more appropriate for person-job fit scenario.


\paratitle{Ablation Study (RQ2)}

The major technical contribution of our approach lies in the dual-perspective interaction graph construction, as well as the involved two optimization objects. 
We now analyze how each part contributes to the final performance.

We consider the following three variants of DPGNN: \textbf{(A)} \underline{DPGNN w/o DPG} replaces the proposed dual-perspective interaction graph to the conventional interaction graph, where each user has only one representation.; \textbf{(B)} \underline{DPGNN w/o QL}
changes quadruple-based loss to BPR loss; \textbf{(C)} \underline{DPGNN w/o SSL} removes the dual-perspective contrastive loss.

In Table~\ref{tab:ablation_study}, we can see that the performance order can be summarized as \underline{DPGNN w/o QL} $<$ \underline{DPGNN w/o DPG} $<$ \underline{DPGNN w/o SSL} $<$ DPGNN. 
These results indicate that all three components are useful to improve the performance of DPGNN.
Especially, the dual-perspective interaction graph and the quadruple-based loss function bring more improvements to our approach.

\begin{figure*}[t!]
	\centering
	\includegraphics[width=1\textwidth]{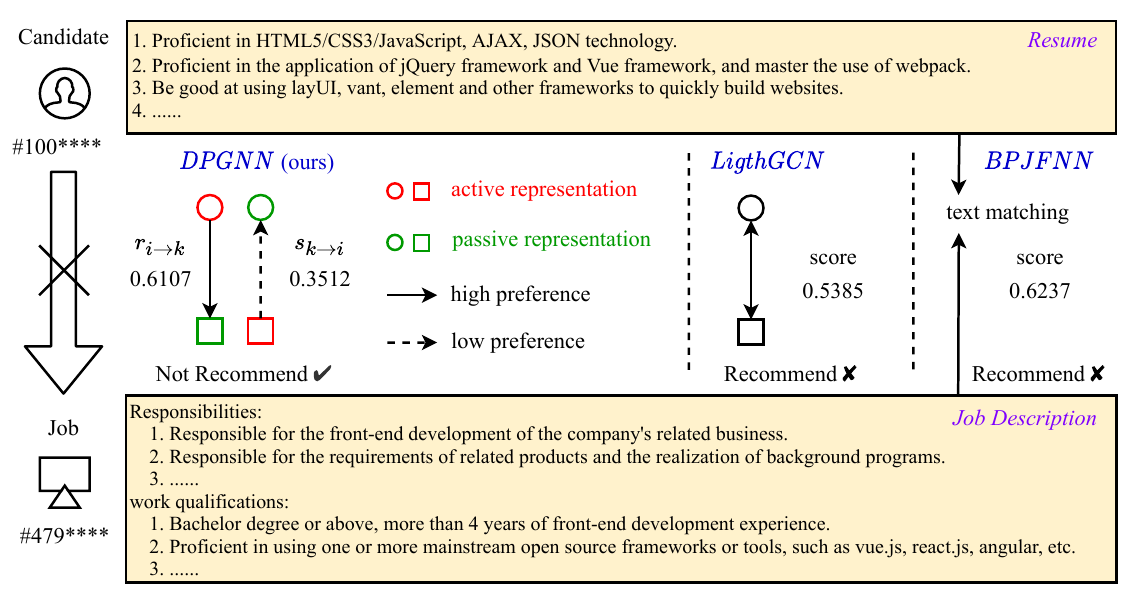}
	\caption{\textcolor{black}{A case where candidate and job have high text similarity but different preferences. Though the candidate prefers the given job position, the recruiter will not give a positive response. 
	This case shows that DPGNN can capture the two-way selection preference, which is helpful to improve the recommendation performance of the person-job fit.
}}
	\label{fig:case_study}
\end{figure*}

\paratitle{Data Sparsity Levels (RQ3)}

As collaborative filtering methods are known to be sensitive to data sparsity, here we give an in-depth analysis to show how the proposed method DPGNN performs on different sparsity levels.
Concretely, we divided all the candidates and jobs into five groups based on the number of their interactions, with the overall number of interactions in each group remaining constant.
Then, we compare the recommendation performance of DPGNN with a strong baseline $\text{LGCN}_{\text{BERT}}$ on these overall ten groups and report the results in Figure~\ref{fig:sparsity}.
We can find that the performance of DPGNN is consistently better than $\text{LGCN}_{\text{BERT}}$, even in those groups where users have sparse interactions.
On the one hand, the preference representations are partially initialized by textual representations encoded. 
In this way, the prediction is not only based on behavior data. 
On the other hand, as contrastive learning has been shown effective on sparse data~\cite{lin2022ncl}, the proposed dual-perspective contrastive learning manner (Eqn.~\eqref{eq:l_cl}) can also benefit recommendation for those users with sparse interactions.
The twofold techniques help DPGNN provide high-quality recommendations to users with different sparsity levels.

\paratitle{Parameter Tuning (RQ4)}

In this part, we examine the robustness of our model and perform detailed analysis of key hyper-parameters. For simplicity, we only incorporate the best baseline $\text{LGCN}_{\text{BERT}}$ from Table~\ref{tab:overall_comparison} as a comparison.

\emph{Varying the layers of GNN.}
We introduce $l$ to denote the layer number of GCN in Eqn~\eqref{eq:gnn_prop}.
Here, we vary the number of layers from $0$ to $4$.
As shown in Figure~\ref{fig:param_tuning}, our model achieves the best performance when $l=3$, indicating DPGNN can deepen the usage of interaction histories.
Overall, the performance is relatively stable when $l\ge1$.
When $l \ge 4$, the model may suffer from the over-smoothing issue and get sub-optimal results.

\emph{Varying the temperature of the contrastive loss.}
The parameter $\tau$ of the contrastive loss is tuned in the ranges of \{0.5, 0.1, 0.05, 0.01, 0.005, 0.001\}.
We can see that with different common values of $\tau$, the performances of DPGNN are constantly better than $\text{LGCN}_{\text{BERT}}$.
Besides, DPGNN achieves the best performance when the temperature is around $0.2$.

\emph{Varying the weight of the contrastive loss.}
The weight of the contrastive loss are tuned in the ranges of \{0.5, 0.3, 0.2, 0.1, 0.07, 0.05, 0.03, 0.01\}.
As shown in Figure~\ref{fig:param_tuning}, for our model, a weight above $0.05$ works better.
The fact that larger $\lambda$ works better indicates the importance of modeling the correlations between preference representations of the same user.

\subsection{Case Study}

As shown in Table~\ref{tab:overall_comparison}, we can see that our approach can achieve better performance than both collaborative filtering methods and content-based methods. 
It is interesting to study that in what situation our method can work better.
For this purpose,
we present an illustrative case in Figure~\ref{fig:case_study}.

We randomly sample a job-candidate pair with their text descriptions from the test set of our experimental dataset Tech.
It can be observed that our approach calculates two opposite one-way selection preference scores $r_{i\rightarrow k}$ and $s_{k \rightarrow i}$, which means that the candidate may like the job but the recruiter doesn't satisfy him (may have better choices).
With these two preference scores, we can see that it is not appropriate to recommend the candidate for the job in this situation. 
But for some content-based (BPJFNN) or collaborative filtering-based (LightGCN) models, the results may be the opposite with a relatively positive score obtained.

This case verifies that person-job fit is a two-way selection process. The preferences from both sides are key factors that determine the final matching. The proposed DPGNN can capture these two-way selection preferences with historical interactions from two sides,
making more effective recommendations.

\section{Conclusion and Future Work}

This paper presented a dual-perspective graph convolution network to model the directed interactions between jobs and candidates for person-job fit.
In our approach, we propose to construct a dual-perspective interaction graph to model the directed behavior. A hybrid preference propagation algorithm is leveraged to learn the node representations.
To optimize the entire model, 
we design an effective algorithm including a quadruple-based loss and a dual-perspective contrastive learning loss and jointly optimize them.
Extensive experiments indicate that the proposed approach can achieve better performance from both perspectives of candidates and job positions compared with competitive baselines.

Besides
textual information and two-way selection behaviors
, there are additional features that are essential for inferring user preferences in real online recruitment platforms, such as timestamps, expected salary ranges, working locations, and chronological sequential behavior histories.
As a result, we consider designing a more unified framework for future work
that incorporates various features into two-way selection preference modeling.
Moreover, when we regard candidates and jobs in the person-job fit scenario as two groups of users, it is related to the group balancing studies in 
multi-stakeholder recommendation~\cite{zheng2021multi}, which focuses on
designing joint optimization objectives towards diversity~\cite{clarke2008novelty, radlinski2009redundancy}, novelty~\cite{ribeiro2013multi} or fairness~\cite{beutel2019fairness, biega2018equity}.
For future work, we will explore how to balance these two stakeholders as well as benefit both groups.

\balance

\begin{acks}
This work was partially supported by National Natural Science Foundation of China under Grant No. 61872369,
Beijing Natural Science Foundation under Grant No. 4222027,  and Beijing Outstanding Young Scientist Program under Grant No. BJJWZYJH012019100020098.
This work was partially supported by Beijing Academy of Artificial Intelligence~(BAAI).
Xin Zhao is the corresponding author.
\end{acks}

\bibliographystyle{ACM-Reference-Format}
\bibliography{ref}

\end{document}